\documentclass[twocolumn,showpacs,amsmath,amssymb,prb]{revtex4}
\usepackage{graphicx}
\usepackage{amssymb}
\usepackage{amsmath}

\begin{document}
\preprint{Elect}
\title{Resistance noise in Bi$_2$Sr$_2$CaCu$_2$O$_{8+\delta}$}

\author{L. Fruchter}
\author{H. Raffy}
\author{Z.Z. Li}

\affiliation{Laboratoire de Physique des Solides, 
Universit\'{e} Paris-Sud, C.N.R.S. UMR 8502, 91405 Orsay cedex, France}

\date{Received: date / Revised version: date}
%
\begin{abstract} 

The resistance noise in a Bi$_2$Sr$_2$CaCu$_2$O$_{8+\delta}$ thin film is found to increase strongly in the underdoped regime. While the increase of the raw resistance noise with decreasing temperature appears to roughly track the previously reported pseudogap temperature for this material, standard noise analysis rather suggests that the additional noise contribution is driven by the proximity of the superconductor-insulator transition.

\end{abstract}

\pacs{72.20.Jv,72.70.+m,74.25.Fy,74.40.+k,74.72.Hs,74.78.Bz}

\maketitle

Since its first observation in YBa$_2$Cu$_3$O$_{7-\delta}$ \cite{alloul1989}, the nature of the pseudogap (PG) phase observed in many underdoped high $T_c$ superconductors is the matter of intense debate. Two distinct points of view still oppose. The PG may originate from quantum disorder above the superconducting transition temperature $T_c$ of the d-wave superconductor, or it is a distinct phase, possibly competing with the superconducting one. There seems to be recent converging evidences that weaken the former view. Indeed, some experiments and theories point toward the fact that the phase disorder mechanism might not be pertinent up to the temperature at which the pseudogap opens\cite{lammert2001,rullier2007,johannsen2007}, while it is now well established that the superconducting and PG phases are associated to two distinct energy gaps\cite{deutscher1999, boyer2007}. At the same time, it is now known that the short range antiferromagnetic correlations can account for both the opening of the pseudogap and its competition with the superconducting order\cite{tremblay2006}. Despite these recent advances, there are still arguments in favor of a pairing gap in the PG phase\cite{hufner2007,feigelman2006} and open questions remain. The problem of electronic inhomogeneity of the high $T_c$ materials has been present since the very beginning, and it was pointed out early that the lightly doped antiferromagnet may experience a phase separation\cite{emery1990}. Since then, it has been observed that the underdoped high $T_c$ material Bi$_2$Sr$_2$CaCu$_2$O$_{8+\delta}$ is subject to electronic phase ordering\cite{vershinin2004,liu2007} and electronic inhomogeneity\cite{cren2000,pan2001,lang2002,kinoda2003,mcelroy2005}. A recent observation for the spatial distribution of the two gaps suggested evidence for some inhomogeneity related to the pseudogap phase only\cite{boyer2007}. 

The presence of dynamic or static domains with different electronic properties may reflect in the transport properties. An attempt in this direction was performed in the work of ref.~\onlinecite{bonetti2004} which reports, for YBa$_2$Cu$_3$O$_{7-\delta}$ nanowires, the occurrence of telegraphlike fluctuations in the resistance, in the pseudogap region. The noise was interpreted as the evidence for the formation of dynamical domain structure. While the investigation of nanowires are advantageous for both the observation of large resistance fluctuations and an estimate of the presumed domain size (at the expend of a possibly large boundary contribution and material degradation), large scale investigations may reflect the same physics - although with a much smaller signature - in the resistance noise. We present here systematic resistance noise measurements performed on Bi$_2$Sr$_2$CaCu$_2$O$_{8+\delta}$ conductors in the 10~$\mu$m range as a function of doping. While at first sight there seems to be for this material also an additional contribution to the transport noise below the pseudogap temperature, standard noise analysis rather suggests that this contribution is restricted to the vicinity of the superconductor-insulator transition.

\section*{}

In order to study the resistance noise as a function of doping, a Bi$_2$Sr$_2$CaCu$_2$O$_{8+\delta}$ thin film, $970$~\AA{} thick, epitaxially grown on a SrTiO$_3$ substrate\cite{li1993}, was patterned in the Wheatstone bridge geometry. DC sputtered gold contacts were deposited on the film. The maximum doping was obtained by annealing in pure oxygen at 420~$^\circ$C and subsequent less doped states were obtained by annealing the sample under vacuum in the temperature range 250$^\circ$C - 270$^\circ$C. Each of the four resistances of the bridge consisted in a 400~$\mu$m x 18~$\mu$m stripe. The low frequency bridge unbalance noise spectrum was measured using the demodulation technique\cite{scofield1987}. It was fed with a 37 $\mu$A AC current at 840 Hz frequency, using a large buffer resistance, so that the contact resistance was negligible. The fluctuating unbalance voltage was applied to the input of a Stanford Research SR830 lock-in, and the sine-convoluted  output was analyzed by a spectrum analyzer. The demodulation allows to reject the 1/f amplifier noise and to benefit at low frequency from its high frequency characteristics. The noise spectrum at each doping state was recorded from the ambient temperature down to 90 K in a liquid nitrogen cryostat. We systematically subtracted the noise recorded in the absence of the AC current from the one with the AC current. This removes the intrinsic Johnson noise, $(4 k_BT R)^{1/2}$ V$_{RMS}$~Hz$^{-1/2}$, which was otherwise at least 5 dB less than the measured sample noise. The background noise resulting from this procedure was about $-180$ dB V$_{RMS}$~Hz$^{-1/2}$.

\begin{figure}
\includegraphics[width= \columnwidth]{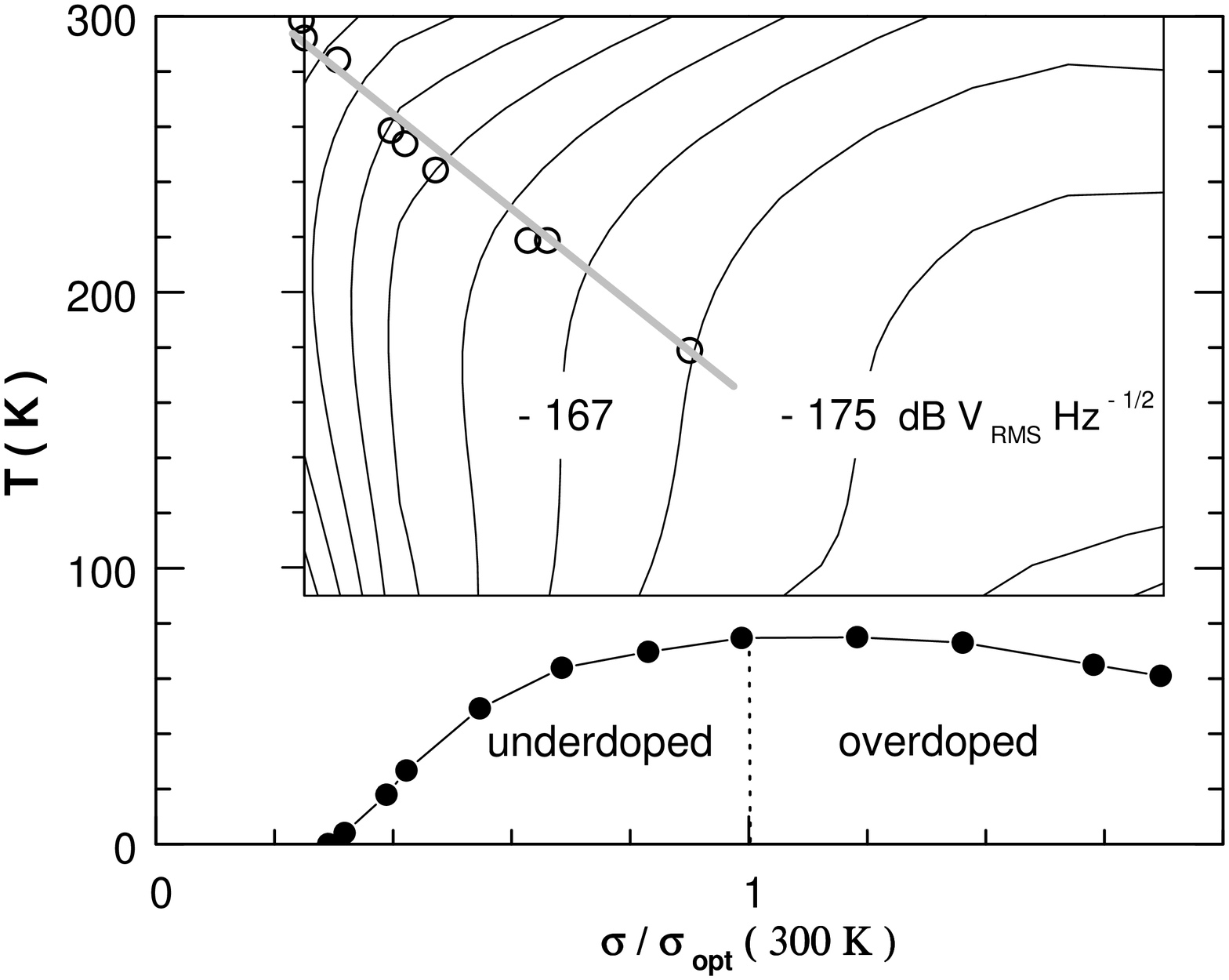}
\caption{Lines : Voltage noise spectrum level at 0.2 Hz for the Wheatstone bridge fed with 37 $\mu$A. Dots delimit the superconducting dome. The conductivity in abscissa, normalized by its value at optimal doping ($\sigma_{opt} = 1.7\,10^5 \,\Omega^{-1} m^{-1}$), is indicative of the doping level. The contour lines (separated by 4 dB V$_{RMS}$ Hz$^{-1/2}$) were obtained from measurements made every 10 K at the doping level corresponding to the dots. Open symbols and line indicate the location for the opening of the pseudogap, according to ref.~\onlinecite{konstantinovic2000}.}\label{contour}
\end{figure}

\begin{figure}
\includegraphics[width= \columnwidth]{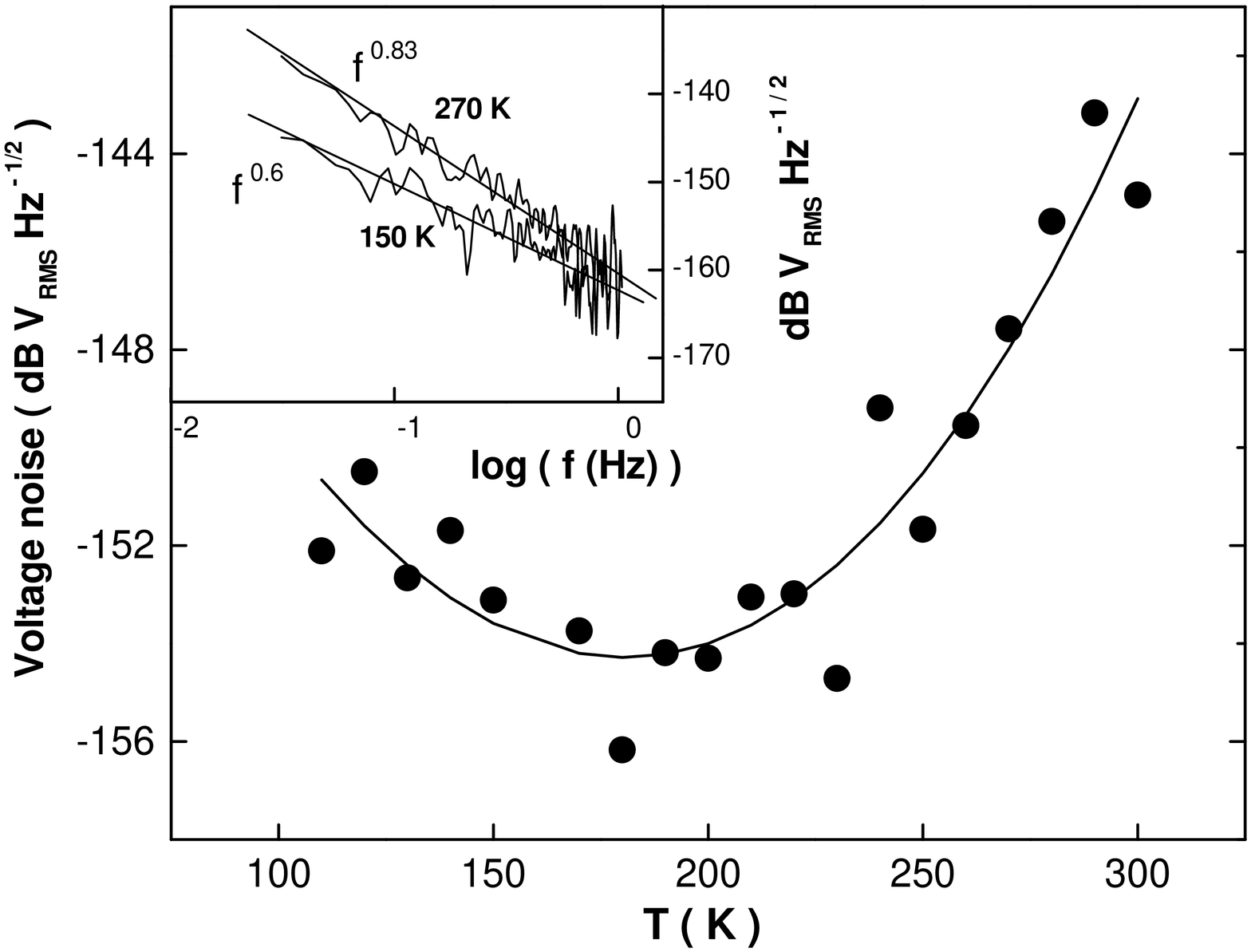}
\caption{Noise spectrum level at 0.2 Hz for the underdoped state with $T_c$ = 18 K, showing re-entrant behavior. The line is a guide to the eye. Inset : noise spectrum, illustrating the power law behavior.}\label{etatC}
\end{figure}

The noise level at frequency $f=0.2$ Hz at the Wheatstone bridge output (equivalent to the one at a single resistance fed by the bridge current) is depicted in Fig.\ref{contour}. Strikingly, not only does the noise level increase strongly when going from the overdoped to the underdoped states, but, also, the pseudogap opening temperature previously obtained for this material\cite{konstantinovic2000} appears as a crossover where the noise level, rather than showing a steady decrease as the temperature is lowered at constant doping, remains constant or - for the most underdoped samples - even shows a re-entrant behavior (Fig.\ref{etatC}). As stated, the Johnson noise cannot account for such a behavior, as it is removed from the measured spectrum (it is also found decreasing with temperature and, for the data in Fig.\ref{etatC}, 10 dB less than the displayed noise density). The understanding of the fact that the noise level increases for the underdoped samples as temperature decreases is undoubtly a key to the comprehension of the crossover depicted in Fig.\ref{contour}, and we discuss this point thereafter.

Our first hypothesis is that there is no change in the electronic properties of the sample as one varies the temperature (such as the occurrence of some phase separation). Although we could not precisely investigate the frequency dependence for the noise level of our sample, due to the rapid merging of the noise signal with the background one for increasing frequency, we find that the power spectrum spectral density is close to the generic $1/f$ behavior in the underdoped regime (Fig.\ref{etatC}). Then, a classical description in terms of thermally activated fluctuators might be appropriate, as it is well known that such 1/f noise is obtained in the case of a uniform distribution of such fluctuators. Within such a model, in order to account for the non monotonic behavior of the noise level, one must admit the existence of two characteristic energies in the fluctuator distribution (both much larger than $k_B\,T$)\cite{dutta1979}. The observation of several distinct activation energies corresponding to oxygen atom jump between double-well potential minima may well be understood in the case of YBa$_2$Cu$_3$O$_{7-\delta}$, as the ordering of the oxygen in the CuO chains plane allows for such distinct sites\cite{bobyl1997}. In the case of Bi$_2$Sr$_2$CaCu$_2$O$_{8+\delta}$, where the mobile oxygen are confined to the BiO layers, it is more difficult to account for these distinct sites and, as long as doping atoms are concerned, one expects a single activation energy for hopping (actually related to the oxygen diffusion constant). The way this can affect transport in the CuO$_2$ plane was discussed in ref.~\onlinecite{mcelroy2005}. Actually, a second hopping mechanism was observed for Bi$_2$Sr$_2$CaCu$_2$O$_{8+\delta}$ from mechanical loss experiments, associated to an activation energy three times less than the one for oxygen diffusion\cite{donzel1995}. The origin of this second contribution is however unclear but, according to ref.~\onlinecite{donzel1995}, the doping dependence of this relaxation peak is similar to that for $T_c$, so that a specific role of this mechanism in the underdoped regime only seems unlikely here.

The second hypothesis is that there is a change in the electronic topology of the hole-doped CuO$_2$ layers. In a general way, on expects, from the Hooge relation\cite{hooge1969},

\begin{equation}
\alpha = N\,S_V\,f / V^2
\end{equation}
 
\begin{figure}
\includegraphics[width= \columnwidth]{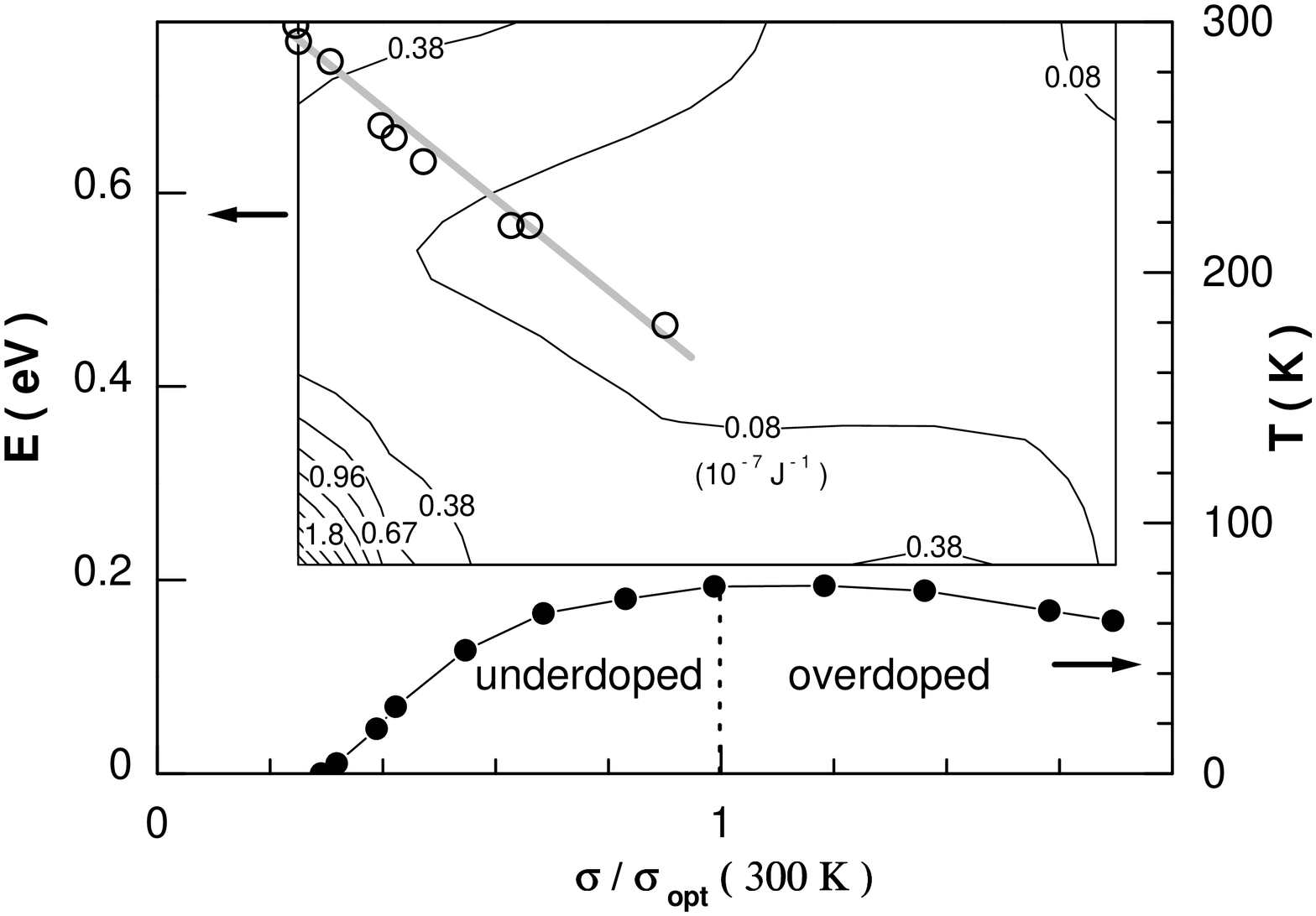}
\caption{Contour plot for the distribution function $\textsl{D}(E)$ obtained from Eq.\ref{equationdistribution}, using $\tau_D = 10^{-13}$ s and $S$(T, f=0.2 Hz) shown in  Fig.~\ref{contour}.}\label{distribution}
\end{figure}

(where $\alpha$ is the Hooge parameter, $V$ is the voltage, $S_V$ is the voltage noise power density and $N$ is the number of carriers in the sample), an increase of the noise spectral density when the fluctuating volume involved in the conductivity decreases (i.e. when $N$ decreases, with $\alpha$ a constant). In particular, this can be the case when the electronic transport is controlled by charge hopping : as was shown in Ref. \onlinecite{kozub1996}, one also expects an \textit{increase} of the noise spectral density as temperature decreases, due to the decrease of the percolation resistor network density as temperature \textit{decreases}, in the case of the Mott-type hopping or hopping within the Coulomb gap. The proximity to a Mott insulator for the strongly underdoped material makes such a mechanism appealing in our case. However, the insulator being quite far away from the optimally doped regime where the additional contribution to the electronic noise starts to manifest itself in the present case, we are left here with a difficulty. Such a difficulty is actually lifted by a proper analysis of the data. In the case of a broad distribution of the activation energy for thermally activated fluctuators involved in the transport noise, this distribution may be obtained in a simple way from the Hooge parameter as :

\begin{equation}
\textsl{D}(E) =  \frac{\alpha(E)}{k_B T N} = \frac{f}{k_BT} \frac{S_V(E)}{V^2}
\label{equationdistribution}
\end{equation}

where $E = - k_B T \ln(2 \pi f \tau_D)$, $\tau_D$ being a characteristic frequency of the order of the Debye frequency. The distribution function $\textsl{D}(E)$ was evaluated in Fig.~\ref{distribution}. 

As can be seen, the distribution obtained in this way is fairly constant and insensitive to doping (with a Hooge parameter at optimal doping $\alpha \approx 10^{-2}$ -- assuming the carrier density $n = 4.8\,10^{21}$ cm$^{-3}$ -- corresponding to a moderately clean metal), and the \textit{apparent} number of defects rises sharply for the most underdoped states only. Therefore, when analyzed in this way, the noise increase does no longer appear as being linked to the opening of the pseudogap, but rather to the proximity of the superconductor-insulator transition. In this case, as stated above, the increase of the noise density as temperature decreases is due to the percolation cluster that shrinks as temperature decreases. Remarkably, this mechanism seems effective for doping states quite far away from the insulating one (Fig.~\ref{distribution}), in a region where the thermally activated conduction mechanism cannot be detected from a straightforward analysis of the resistivity : this corresponds to the `anomalous insulator' in ref.~\onlinecite{ando1995} (see also Ref.~\onlinecite{lai1998} for further references and discussion), where suppression of superconductivity reveals the insulating behavior. Beloborodov et al\cite{beloborodov2007} have suggested that the anomalous insulator may be identified -- in the case of La$_{2-x}$Sr$_x$CuO$_4$ -- as a granular metal due to a phase segregation throughout the underdoped regime, as could be the case also for Bi$_2$Sr$_2$CaCu$_2$O$_{8+\delta}$. Within this picture, the fluctuating bonds at the origin of the transport noise could influence the conduction between the grains.

Let us now compare our results with the ones in Ref.~\onlinecite{bonetti2004}. In Ref.~\onlinecite{bonetti2004}, the noise for underdoped YBa$_2$Cu$_3$O$_{7-\delta}$ has been resolved down to a single fluctuator, which, according to the authors, could be the switching of a single stripe domain. The typical switching time for such a domain - assumed to fluctuate between a highly resistive state and the bulk value - at 100 K is of the order of 1 s. The typical defect domain would be set by the correlation length for a stripe, $\Lambda \approx$ 40 nm, and the defect concentration was estimated c = 1\%. For a frequency of the order of the switching time, $\tau$, we expect for the noise spectral density $S_V \approx V^2 \tau c / N$, where $N$ is the number of domains, that is in our case $N \simeq 3\,10^8$ and, using $V \simeq 40$ mV as in the present study, one has $S_V \approx 10^{-14}$ V$^2$ Hz$^{-1}$. This is 3 orders of magnitude larger than typically observed here for underdoped states. At the same time, the Hooge parameter for YBa$_2$Cu$_3$O$_{7-\delta}$ thin films grown on LaAlO$_3$ substrates, as in Ref.~\onlinecite{bonetti2004}, was found in Ref.~\onlinecite{liu1994} also 3 orders of magnitude larger than for the present study (see also Ref. \onlinecite{bobyl1997}). This probably illustrates the fact that epitaxy on such a substrate -- which is heavily twinned -- induces a large amount of defects in the grown material. Such a contribution might have been overlooked in Ref. \onlinecite{bonetti2004}. Of course, we cannot exclude that charge ordering anisotropy is an essential ingredient for the unstabilities observed in Ref. \onlinecite{bonetti2004}, and that it is missing\cite{vershinin2004,liu2007} in the case of Bi$_2$Sr$_2$CaCu$_2$O$_{8+\delta}$, nor that the domain parameters are orders of magnitude different for these materials. However, classical fluctuators as evidenced in Refs.~\onlinecite{liu1994} and ~\onlinecite{bobyl1997} are likely to show up in nanoscaled resistors also, and sorting such a contribution from the proposed exotic mechanism for fluctuations should be a precondition to the full understanding of the resistance noise in these materials.

In conclusion, we have found that the resistance noise in a Bi$_2$Sr$_2$CaCu$_2$O$_{8+\delta}$ thin film likely arises due to the proximity of the superconductor - insulator transition at low doping. We found no clear evidence for an additional contribution, that would be due to the existence of fluctuating domains specific to the pseudogap state.

\begin{acknowledgments}
We thank F. Bouquet for discussions and critical reading of this manuscript, and H. Bouchiat for useful advices for noise measurement techniques.
\end{acknowledgments}


\begin{thebibliography}{}

\bibitem{alloul1989}
H. Alloul, T. Ohno and P. Mendels, Phys. Rev. Lett. \textbf{63}, 1700 (1989).

\bibitem{lammert2001}
Paul E. Lammert and Daniel S. Rokhsar, cond-mat/0108146 (2001).

\bibitem{rullier2007}
F. Rullier-Albenque, H. Alloul, Cyril Proust, P. Lejay, A. Forget and D. Colson, Phys. Rev. Lett. \textbf{99}, 027003 (2007).

\bibitem{johannsen2007}
N. Johannsen, Th. Wolf, A. V. Sologubenko, T. Lorenz, A. Freimuth and J. A. Mydosh, cond-mat/0706.1463 (2007).

\bibitem{deutscher1999}
G. Deutscher, Nature \textbf{397}, 410 (1999).

\bibitem{boyer2007}
M. C. Boyer, W. D. Wise, Kamalesh Chatterjee, Ming Yi, Takeshi Kondo, T. Takeuchi, H. Ikuta and E. W. Hudson, cond-mat/0705.1731 (2007).

\bibitem{tremblay2006}
A.-M.S. Tremblay, B. Kyung and D. S\'en\'echal, Low Temp. Phys. \textbf{32}, 424 (2006).

\bibitem{hufner2007}
S. H\"ufner, M.A. Hossain, A. Damascelli and G.A. Sawatzky, cond-mat/0706.4282 (2007).

\bibitem{feigelman2006}
M. V. Feigel'man, L. B. Ioffe, V. E. Kravtsov and E. A. Yuzbashyan, Phys. Rev. Lett. \textbf{98}, 027001 (2007).

\bibitem{emery1990}
V. J. Emery, S. A. Kivelson, and H. Q. Lin, Phys. Rev. Lett. \textbf{64}, 475 (1990).

\bibitem{vershinin2004}
M. Vershinin et al, Science \textbf{303}, 1995 (2004).

\bibitem{liu2007}
Y. H. Liu, K. Takeyama, T. Kurosawa, N. Momono, M. Oda and M. Ido, Phys. Rev. B \textbf{75}, 212507 (2007).

\bibitem{cren2000}
T. Cren, D. Roditchev, W. Sacks, J. Klein, J.-B. Moussy, C. Deville-Cavellin and M. Lag\"ues, Phys. Rev. Lett. \textbf{84}, 147 (2000).

\bibitem{pan2001}
S.H. Pan, J.P. O'Neal, R.L. Badzey, C. Chamon, H. Ding, J.R. Engelbrecht, Z. Wang, H. Eisaki, S. Uchida, A.K. Gupta, K.W. Ng, E.W. Hudson, K.M. Lang, J.C. Davis, Nature \textbf{413}, 282 (2001).

\bibitem{lang2002}
K. M. Lang, V. Madhavan, J. E. Hoffman, E. W. Hudson, H. Eisaki, S. Uchida and J. C. Davis,
Nature \textbf{415}, 412 (2002).

\bibitem{mcelroy2005}
K. McElroy, Jinho Lee, J.A. Slezak, D.-H. Lee, H. Eisaki, S. Ushida and J.C. Davis, Sience \textbf{309}, 1048 (2005).

\bibitem{kinoda2003}
G. Kinoda, T. Hasegawa, S. Nakao, T. Hanaguri, K. Kitazawa, K. Shimizu, J. Shimoyama and K. Kishio, Phys. Rev. B \textbf{67}, 224509 (2003).

\bibitem{bonetti2004}
J. A. Bonetti, D. S. Caplan, D. J. Van Harlingen and M. B.Weissman, Phys. Rev. Lett. \textbf{93}, 087002 (2004).

\bibitem{li1993}
Z.Z. Li, H. Rifi, A. Vaures, S. Megtert and H. Raffy, Physica C \textbf{206}, 367 (1993).

\bibitem{scofield1987}
J. H. Scofield, Rev. Sci. Inst. \textbf{58}, 985 (1987).

\bibitem{konstantinovic2000}
Z. Konstantinovic, Z.Z. Li and H. Raffy, Physica C \textbf {341-348}, 859 (2000).

\bibitem{dutta1979}
P. Dutta, P. Dimon and P.M. Horn, Phys. Rev. Lett. \textbf{43}, 646 (1979).

\bibitem{bobyl1997}
A.V. Bobyl, M.E. Gaevski, S.F. Karmanenko, R.N. Kutt, R.A. Suris, I.A. Khrebtov, A.D. Tkachenko and A.I. Morosov, J. Appl. Phys. \textbf{82}, 1274 (1997).

\bibitem{donzel1995}
L. Donzel, Y. Mi and R. Schaller, Physica C \textbf{250}, 75 (1995).

\bibitem{hooge1969}
F.N. Hooge, Phys. Lett. \textbf{29A}, 139 (1969).

\bibitem{kozub1996}
V.I. Kozub, Solid State Com. \textbf{97}, 843 (1996).

\bibitem{ando1995}
Y. Ando, G. S. Boebinger, A. Passner, Tsuyoshi Kimura, and Kohji Kishio, Phys. Rev. Lett. \textbf{75}, 4662 (1995).

\bibitem{lai1998}
E. Lai and R.J. Gooding, Phys. Rev. B \textbf{57}, 1498 (1998).

\bibitem{beloborodov2007}
I. S. Beloborodov, A. V. Lopatin, V. M. Vinokur and K. B. Efetov, Rev. Mod. Phys. \textbf{79}, 469 (2007).

\bibitem{liu1994}
Li Liu, K. Zhang, H.M. Jaeger, D.B. Buchholz and R.P.H. Chang, Phys. Rev. B \textbf{49}, 3679 (1994).


\end{thebibliography}
\end{document}